\documentclass[aps,showpacs,showkeys,preprint]{revtex4}
\usepackage[dvips]{graphicx}

\begin{document}
\preprint{KOBE-TH-02-02}

\title{How precisely can we reduce the three-flavor neutrino oscillation
to the two-flavor one only from $\frac{\delta m^{2}_{12}}{\delta m^{2}_{13}}\lesssim
\frac{1}{15}$ ?}


\author{~C.~S.~Lim}
\email[]{lim@phys.sci.kobe-u.ac.jp}
\author{~K.~Ogure}
\email[]{ogure@phys.sci.kobe-u.ac.jp}
\author{~H.~Tsujimoto}
\email[]{tujimoto@phys.sci.kobe-u.ac.jp}
\affiliation{Department of Physics, Kobe University, Rokkodaicho 1-1,
Nada ward, Kobe 657-8501, Japan}


\date{\today}

\begin{abstract}
We derive a reduction formula that expresses the survival rate for
the three-flavor neutrino oscillation using the two-flavor one to next-to-leading order when there is one resonance due to the matter
effect.  We numerically find that the next-to-leading reduction formula is extremely accurate and the improvement is relevant for the precision test of solar neutrino oscillation and the indirect measurment of CP violation in the leptonic sector.  We also derive a
reduction formula, which is slightly different from that previously
obtained, in the case when there are two resonances.  We
numerically verify that this reduction formula is quite accurate and is valid for a wider parameter region than are those previously obtained.
\end{abstract}



\maketitle

\section{Introduction\label{intro}}

The physics of neutrino oscillation is currently under very active
investigation, since it leads to physics beyond the standard model.
Two-flavor neutrino oscillation, however, is adopted in most analyses of
the data, although everyone knows that there are three active neutrino
flavors.  Two-flavor oscillation is easy to investigate in
comparison with three-flavor oscillation because there are only two
parameters: a mass-squared difference and a mixing angle.  In addition, some
exact solutions of the oscillation probability are known in the
two-flavor oscillation scheme even in the presence of a matter
effect \cite{Hax,Par,Tos,Pet,Kan,Kuo,Not,Fri}.  To make the analysis
realistic, however, we need to work in the full three-flavor system,
where we have six parameters, two mass-squared differences,  three mixing angles,
and one CP phase.  The analysis using three-flavor oscillation is
particularly important when we compare the outputs from different
experiments, sensitive to different mass-squared differences
and mixing angles. 
The importance of the analysis in the full three-flavor context for terrestrial neutrino oscillation experiments has been discussed \cite{Shrock,Gon2}.

The simplest way to investigate the three-flavor oscillation is by relying on a numerical calculation.  There is no serious technical
difference between the two-flavor oscillation and the three-flavor one
in numerical calculations \footnote{ There is, however, a trick only for
the two-flavor neutrino oscillation.  The equation of motion of the
two-flavor neutrino oscillation, which is a coupled first order
differential equation, can be converted to a single {\it first order}
differential equation\cite{Fri}.  This makes the numerical calculation
quite easy.}.  The parameter space, however, becomes very large, six
dimensional, and is difficult to exhaust.  What is worse, we
cannot easily understand the physical consequences from the numerical
results intuitively, even if the parameter space is exhausted.

A more elegant way to investigate the three-flavor oscillation is to
reduce the three-flavor oscillation to an effective two-flavor one
using the hierarchy between the two mass differences or the smallness of
some mixing angle.  For example, the following reduction formula for the
survival probability of the electron neutrino, $S_{3\nu}$ is found in
the case that one mass difference is much larger than the other one
and the matter effect, $\Delta_{12},A(x)\ll \Delta_{13}$ \cite{Lim},
\begin{eqnarray}
S_{3\nu}=S_{2\nu,eff}\cos^4{\theta_{13}}+\sin^4{\theta_{13}}.
\label{leading}
\end{eqnarray}
Here, $\theta_{13}$ is the mixing angle defined in Sec. \ref{onenext},
while $S_{2\nu,eff}$ is the survival rate calculated in the effective
two-flavor scheme with effective matter effect $A_{eff}(x)\equiv
\cos^2{\theta_{13}}A(x)$.  We show the definitions of each quantity in
detail in Sec. \ref{one}.  This relation is often used in the
analysis of the solar
neutrino \cite{Shi,Smi1,Fog1,Smi2,Fog2,Fog3,Nun,Gon}.  Although it is
exact in the limit $\Delta_{12}/\Delta_{13},
A(x)/\Delta_{13}\rightarrow 0$, the actual hierarchy is not so
good.  Taking the large mixing angle Mikheyev-Smirnov-Wolfenstein (LMA-MSW) solution for the solar neutrino problem and
the allowed range of $\Delta _{13}$ from the analysis of the atmospheric
neutrino oscillation, we find that the hierarchies are rather mild,
$\Delta_{12}/\Delta_{13}\lesssim 1/15$,
$A/\Delta_{13}\lesssim 1/10$ \footnote{ The atmospheric and
reactor data also indicate that the hierarchy of the mass differences
is not so large\cite{Gon2}.  }.  This indicates that the above
relation possibly has an error around several percents or more.

Such an error is not so important in just verifying the existence of the
neutrino masses.  Current interest in neutrino physics is,
however, not only in verifying the finite masses but also in determining
the precise values of the parameters.  The allowed region of the mass
differences and the mixing angles may be affected due to error in
the formula used in the analysis.  This error would be more serious in
attempts to observe the CP violation in the leptonic sector using the
sizes of the unitarity triangle, since it requires a more precise
determination of the survival rate \cite{Joe,Xin,Bra,Smi3}. 

One of the aims of the present paper is to examine how precise the
reduction formula is.  We then propose a next-to-leading order
reduction formula, which is surprisingly precise.  We assume only the
mass hierarchy $\Delta_{12}/\Delta_{13}\lesssim 1/15$ and do
not impose any restriction on $\theta_{13}$ in order to keep our
analysis as general as possible.  While we average the survival rate
with respect to the final time, we do not average it with respect to
the initial time.  The reduction formula obtained therefore can be
used without averaging over the initial time.

Another example of the reduction formula is the following relation for
 two successive resonances \cite{Kuo}:
\begin{eqnarray}
S_{3\nu}
=
P_L P_H\cos^2{\theta_{12}}\cos^2{\theta_{13}}
+P_H(1-P_L)\sin^2{\theta_{12}}\cos^2{\theta_{13}}
+(1-P_H)\sin^2{\theta_{13}}.
\label{leading2}
\end{eqnarray}
Here, $\theta_{12}$ is the mixing angle defined in Sec. \ref{onenext}, and $P_L$ and $P_H$ are the jump probabilities for
resonances with lower and higher number densities, respectively.
This relation is used for the investigation into neutrino
oscillation inside a supernova, which has two successive resonances
due to the high matter density \cite{Kuo2,Dut,Smi4,Fog4,Fog5}.  It is
also used for investigation into hypothetical O(GeV) neutrinos from the sun
produced by the annihilation of weakly interacting massive particles (WIMPs), which have two successive
resonances due to the higher neutrino energy \cite{Gou}.  In
Ref. \cite{Gou}, an improved reduction formula Eq. (\ref{Gred}) is derived. 
The reduction formulas Eq. (\ref{leading2}) and  Eq. (\ref{Gred}), are
expressed by the jump probability, while the reduction formula
Eq. (\ref{leading}) is expressed by the survival rate.  We express the
reduction formula by the survival rate also in this case and numerically
study the difference between these expressions.

The present paper is organized as follows.  In Sec. \ref{onenext}, we
derive the next-to-leading order reduction formula in the case of
one resonance with the matter effect. The main result in this section is
the next-to-leading reduction formula Eq. (\ref{final}). In
Sec. \ref{onenum}, we examine the validity of the next-to-leading order
reduction formula using numerical calculation.  In Sec. \ref{twoder},
we derive the reduction formula in the case of two resonances.  The
reduction formula Eq. (\ref{twofinal}), which is the main result of this
section, is expressed without the jump probability.  In
Sec. \ref{twonum}, we numerically verify the validity of the reduction
formula obtained in Sec. \ref{twoder}, comparing to that expressed by
the jump probability.  In Sec. \ref{sum}, we summarize the results
obtained in the present paper.

\section{Reduction formula for one resonance\label{one}}

We derive the next-to-leading order reduction formula in the case of
one resonance in this section.  While the formula is suitable for
investigation of the solar neutrino problem, it would be applicable for other
cases.  We then verify its validity using numerical calculation.

\subsection{Derivation of the next-to-leading order reduction formula
\label{onenext} }
We derive the next-to-leading order reduction formula in the
presence of one resonance. Neutrino
propagation in matter for three flavors is governed by
\begin{eqnarray}
i\frac{d\phi}{dt}
=
\left\{
V
\left(
\begin{array}{ccc}
0 &  &  \\
 & \Delta_{12} &  \\
 &  & \Delta_{13} \\
\end{array}
\right)
V^{\dagger}
+
\left(
\begin{array}{ccc}
A(t) &  &  \\
 & 0 &  \\
 &  & 0 \\
\end{array}
\right)
\right\}
\phi
\label{prope}
\end{eqnarray}
in the base of the weak eigenstate $\phi_{\alpha}\ (\alpha = e,\mu
,\tau)$.  Here, $A(t)=\sqrt{2}G_F N_e(t)$ is the matter induced mass
of the electron neutrino with $N_{e}(t)$ being the electron number
density. The quantities $\Delta_{12}\equiv\frac{\delta m_{12}^2(=m_2^2-m_1^2)}{2 E}$
and $\Delta_{13}\equiv\frac{\delta m_{13}^2(=m_3^2-m_1^2)}{2 E}$, are given in terms of
the neutrino masses $m_i\ (i=1,2,3)$ and the neutrino energy $E$.
Throughout the present paper, we assume a hierarchical relation
$0<\Delta_{12}\ll\Delta_{13}$.  The mixing matrix $V$ is parametrized
as
\begin{eqnarray}
V
&=&
\left(
\begin{array}{ccc}
c_{12} c_{13} & s_{12} c_{13}  & s_{13}e^{-i\delta}  \\
-s_{12} c_{23} -c_{12} s_{23} s_{13} e^{i\delta}
& c_{12} c_{23} -s_{12} s_{23} s_{13} e^{i\delta}&s_{23}c_{13} \\
s_{12} s_{23} -c_{12} c_{23} s_{13} e^{i\delta} 
& -c_{12} s_{23} -s_{12} c_{23} s_{13} e^{i\delta}&c_{23}c_{13}  \\
\end{array}
\right)\\
&=&
\left(
\begin{array}{ccc}
1 & 0  & 0 \\
0 & c_{23} & s_{23} \\
0 & -s_{23} & c_{23} \\
\end{array}
\right)
\left(
\begin{array}{ccc}
c_{13} & 0  & s_{13}e^{-i\delta} \\
0 & 1 & 0 \\
-s_{13}e^{i\delta} & 0 & c_{13} \\
\end{array}
\right)
\left(
\begin{array}{ccc}
c_{12} & s_{12}  & 0 \\
-s_{12} & c_{12} & 0 \\
0 & 0 & 1 \\
\end{array}
\right)
\equiv
V_{23}V_{13}V_{12},
\label{mixmat}
\end{eqnarray}
where $s_{ij}$ and $c_{ij}$ represent $\sin{\theta_{ij}}$ and
$\cos{\theta_{ij}}$, respectively. Since we discuss only the survival
rate of the electron neutrino, which does not depend on $\theta_{23}$
and $\delta$, we set them to zero in the present paper.

For our purpose, it is convenient to work in the base $\xi\equiv
V_{13}^{\dagger} V_{23}^{\dagger} \phi$, where the time evolution
equation (\ref{prope}) is rewritten as
\begin{eqnarray}
i\frac{d\xi}{dt}
&=&
\left\{
V_{12}
\left(
\begin{array}{ccc}
0 &  &  \\
 & \Delta_{12} &  \\
 &  & \Delta_{13} \\
\end{array}
\right)
V^{\dagger}_{12}
+
V_{13}^{\dagger}
\left(
\begin{array}{ccc}
A(t) &  &  \\
 & 0 &  \\
 &  & 0 \\
\end{array}
\right)
V_{13}
\right\}
\xi\\
&=&
\left\{
\left(
\begin{array}{ccc}
\Delta_{12} s_{12}^2 +A c_{13}^2 & \Delta_{12} s_{12} c_{12} & 0 \\
\Delta_{12} s_{12} c_{12} & \Delta_{12} c_{12}^2 & 0 \\
0 & 0 & \Delta_{13}+ A s_{13}^2 \\
\end{array}
\right)
+
\left(
\begin{array}{ccc}
0 & 0 &A s_{13} c_{13}  \\
0 & 0 & 0 \\
A s_{13} c_{13} & 0 & 0 \\
\end{array}
\right)
\right\}
\xi.
\label{propxi}
\end{eqnarray}
Under the hierarchy $\Delta_{12}, A(x)\ll \Delta_{13}$, the second matrix in the right-hand side (RHS) may be treated as a perturbation, while the 2$\times $2 submatrix in the first matrix should not be dealt with as a small perturbation, because neglect of the submatrix causes a degeneracy of eigenvalues. Thus, neglecting the second matrix, we obtain a block diagonalized equation, and
the leading order reduction formula Eq. (\ref{leading}) is easily obtained.  In this
approximation, the discarded terms have the magnitude of order
$O(As_{13} c_{13})$, which handles the  error of this formula.  We then derive a next-to-leading order reduction formula to reduce the magnitude of the error.

For this purpose, let us rewrite Eq. (\ref{propxi}) as
\begin{eqnarray}
i\frac{d\xi}{dt}
&=&
\left\{
\left(
\begin{array}{ccc}
0 & \Delta_{12} s_{12} c_{12} & 0 \\
\Delta_{12} s_{12} c_{12} & \Delta_{12} c_{12}^2 & 0 \\
0 & 0 & 0 \\
\end{array}
\right)
+
\left(
\begin{array}{ccc}
\Delta_{12} s_{12}^2 +A c_{13}^2& 0 &A s_{13} c_{13}  \\
0 & 0 & 0 \\
A s_{13} c_{13} & 0 & \Delta_{13}+A s_{13}^2 \\
\end{array}
\right)
\right\}
\xi.
\label{propxi2}
\end{eqnarray}
In order to incorporate the effect of the perturbation $As_{13}c_{13}$,
we then diagonalize the second matrix by moving to a new base
$\eta=V_{\epsilon}^{\dagger}\xi$, with $V_{\epsilon}$ being a time
dependent unitary matrix,
\begin{eqnarray}
V_{\epsilon}
&=&
\left(
\begin{array}{ccc}
\cos{\epsilon (t)} & 0  & \sin{\epsilon (t)}  \\
0 &  1 & 0 \\
-\sin{\epsilon (t)} & 0 &\cos{\epsilon (t)} \\
\end{array}
\right).
\end{eqnarray}
The diagonalization goes as
\begin{eqnarray}
V_{\epsilon}^{\dagger}
\left(
\begin{array}{ccc}
\Delta_{12} s_{12}^2 +A c_{13}^2& 0 &A s_{13} c_{13}  \\
0 & 0 & 0 \\
A s_{13} c_{13} & 0 & \Delta_{13}+A s_{13}^2 \\
\end{array}
\right)
V_{\epsilon}
=
\left(
\begin{array}{ccc}
k_-& 0 & 0  \\
0 & 0 & 0 \\
0 & 0 & k_+ \\
\end{array}
\right),
\end{eqnarray}
where the two eigenvalues are given as
\begin{eqnarray}
k_{\pm}
=
\frac{1}{2}
\left[
A+\Delta_{13}+s_{12}^2\Delta_{12}
\pm
\sqrt{
(A+\Delta_{13}+s_{12}^2\Delta_{12})^2
-4\{\Delta_{13}\Delta_{12}s_{12}^2+A(c_{13}^2\Delta_{13}+s_{13}^2 s_{12}^2 \Delta_{12}) \}
}
\right],\nonumber\\ 
\label{kvalue}
\end{eqnarray}
and the angle $\epsilon$ satisfies the relation
\begin{eqnarray}
\tan{2 \epsilon (t)}
=
\frac{A(t) \sin{2\theta_{13}}}{\Delta_{13}-s_{12}^2 \Delta_{12}-A(t) \cos{2 \theta_{13}}}.
\end{eqnarray}
Although the hierarchy $\Delta_{12}, A\ll \Delta_{13}$ implies that the
time dependent angle is small,
\begin{eqnarray}
\epsilon
\simeq
\frac{A \sin{2 \theta_{13}}}{2 \Delta_{13}},
\label{epsilon}
\end{eqnarray}
the angle plays an important role in the improvement of the reduction
formula.  While the second matrix on the RHS of Eq. (\ref{propxi2})
is diagonalized, we have additional terms in the time evolution
equation in the base of $\eta $. First, since the base $\eta $ depends
on time, the left-hand side of Eq. (\ref{propxi2}) yields the following
extra contribution:
\begin{eqnarray}
-iV_{\epsilon}^{\dagger}
\frac{d V_{\epsilon}}{dt}
=
i\epsilon ^{'}(t)
\left(
\begin{array}{ccc}
0& 0 & -1  \\
0 & 0 & 0 \\
1 & 0 & 0 \\
\end{array}
\right).
\end{eqnarray}
This term has a magnitude of the following order:
\begin{eqnarray}
\left|\epsilon ^{'}(t)\right|
\simeq
\left|\frac{A^{'} \sin{2 \theta_{13}}}{2 \Delta_{13}}\right|
\sim
\frac{A \sin{2 \theta_{13}}}{2 \Delta_{13} L},
\end{eqnarray}
where $L$ denotes the typical length in which the matter effect $A$
changes. In the case of the sun, the typical length corresponds to the
scale height, $L=R_s\sim \frac{R_{\odot}}{10.54}$ \cite{Bah}.

Second, the first matrix in Eq. (\ref{propxi2}) is also slightly
altered due to the change of base to
\begin{eqnarray}
V_{\epsilon}^{\dagger}
\left(
\begin{array}{ccc}
0 & \Delta_{12} s_{12} c_{12} & 0 \\
\Delta_{12} s_{12} c_{12} & \Delta_{12} c_{12}^2 & 0 \\
0 & 0 & 0 \\
\end{array}
\right)
V_{\epsilon}
&=&
\Delta_{12}
\left(
\begin{array}{ccc}
0 & s_{12} c_{12} c_{\epsilon}& 0  \\
s_{12} c_{12} c_{\epsilon} & c_{12}^2 & s_{12} c_{12} s_{\epsilon} \\
0 & s_{12} c_{12} s_{\epsilon} & 0 \\
\end{array}
\right)\\
&=&
\Delta_{12}
\left(
\begin{array}{ccc}
0 & s_{12}c_{12}(1 + O(\epsilon^2))& 0  \\
s_{12}c_{12}(1 + O(\epsilon^2)) & c_{12}^2 &s_{12}c_{12}O(\epsilon)  \\
0 &  s_{12}c_{12}O(\epsilon) & 0 \\
\end{array}
\right).\nonumber\\
\end{eqnarray}

Ignoring the small extra off-diagonal terms, we approximately obtain the propagation equation in the time dependent base $\eta$:
\begin{eqnarray}
i\frac{d\eta}{dt}
&=&
\left(
\begin{array}{ccc}
k_{-}&\Delta_{12}s_{12}c_{12}+O(\Delta_{12}\sin{2\theta_{12}}\epsilon^2)
&O(\frac{A\sin{2\theta_{13}}}{L\Delta_{13}}) \\
\Delta_{12}s_{12}c_{12}+O(\Delta_{12}\sin{2\theta_{12}}\epsilon^2)
&\Delta_{12}c_{12}^2
&O(\Delta_{12}\sin{2\theta_{12}}\epsilon)\\
O(\frac{A\sin{2\theta_{13}}}{L\Delta_{13}}) 
&  O(\Delta_{12}\sin{2\theta_{12}}\epsilon) & k_{+} \\
\end{array}
\right)
\eta\nonumber\\
&\simeq&
\left(
\begin{array}{ccc}
{\tilde A}_{eff} +\Delta_{12}s_{12}^2 &\Delta_{12}s_{12}c_{12}& 0 \\
\Delta_{12}s_{12} c_{12} & \Delta_{12}c_{12}^2 &0\\
0 & 0  & k_{+} \\
\end{array}
\right)
\eta,
\label{propeta}
\end{eqnarray}
where
\begin{eqnarray}
{\tilde A}_{eff}
\equiv
k_- -\Delta_{12}s_{12}^2
.
\label{aeffdef}
\end{eqnarray}
This effective matter effect is well approximated under the hierarchy
$\Delta_{12},A(x)\ll \Delta_{13}$ as
\begin{eqnarray}
{\tilde A}_{eff}
\rightarrow
c_{13}^2A-\frac{\sin^2{2\theta_{13}}A^2}{4\Delta_{13}}
,
\label{aefflim}
\end{eqnarray}
and reduces to $A_{eff}$ used in the leading formula in the limit
$\frac{A}{\Delta_{13}}\rightarrow 0$.  The off-diagonal terms
discarded in Eq. (\ref{propeta}) are smaller than the second matrix of
Eq. (\ref{propxi}) by the order
$\frac{\Delta_{12}\sin{2\theta_{12}}}{\Delta_{13}}$ in the case
\begin{eqnarray}
\frac{A\sin {2\theta_{13}}}{L\Delta_{13}}
\lesssim
\Delta_{12}\sin{2\theta_{12}}\epsilon,
\label{cond}
\end{eqnarray}
which can be rewritten as
\begin{eqnarray}
E
\lesssim
L\delta m_{12}^{2} \sin{2\theta_{12}}.
\label{cond2}
\end{eqnarray}
This condition is well satisfied for the solar neutrinos, since the parameters are $E\lesssim
10 [\textrm{MeV}], L=R_{s}\simeq 3\times 10^{14} [\textrm{eV}^{-1}]$,
and $\delta m^{2}_{12}\sim 10^{-4} [\textrm{eV}^{2}]$. We also discard
the element of $O(\Delta_{12}\sin{2\theta_{12}}\epsilon^2)$ in the
$2\times 2$ submatrix because of its smallness.
Thus, the order of the neglected matrix elements in Eq. (\ref{propeta}) is much reduced, compared with that in the leading order formula, and a considerable improvement is expected in the next-to-leading order formula obtained below.

The survival rate for the electron neutrino is calculated, by solving Eq. (\ref{propeta}), to be
\begin{eqnarray}
S(\nu_e \rightarrow \nu_e)
&=&
\left|
\left(
\begin{array}{ccc}
1& 0& 0
\end{array}
\right)
V_{23} V_{13} V_{\epsilon}(t)
\left(
\begin{array}{cc}
{\bf R}_{2\times 2}(t,0) &{\bf 0}\\
{\bf 0}^{T} & \exp({-i\int_0^{t} k_+ dt})\\
\end{array}
\right)
{V_{\epsilon}(0)}^{\dagger}
V_{13}^{\dagger}
V_{23}^{\dagger}
\left(
\begin{array}{c}
1\\
0\\
0\\
\end{array}
\right)
\right|^2\nonumber\\
&=&
\cos^{2}{\theta_{13}}\cos^2({\theta_{13}+\epsilon(0)})
\left|
\left(
\begin{array}{cc}
1& 0
\end{array}
\right)
{\bf R}_{2\times 2}(t,0)
\left(
\begin{array}{c}
1\\
0\\
\end{array}
\right)
\right|^2
+
\sin^{2}{\theta_{13}}\sin^2({\theta_{13}+\epsilon(0)}).\nonumber\\
\end{eqnarray}
Here, ${\bf R}_{2\times 2}(t,0)$ is the resolvent matrix for the 
two-flavor neutrino oscillation with the matter effect ${\tilde
A}_{eff}$ and mass squared difference $\Delta_{12}$. Thus we get the next-to-leading order reduction formula
\begin{eqnarray}
S_{3\nu}(\nu_e \rightarrow \nu_e)
=
\cos^{2}{\theta_{13}}\cos^2({\theta_{13}+\epsilon(0)})
{\tilde S}_{2\nu ,eff}({\tilde A}_{eff},\Delta_{12})
+
\sin^{2}{\theta_{13}}\sin^2({\theta_{13}+\epsilon(0)}).
\label{final}
\end{eqnarray}
Here, ${\tilde S}_{2\nu ,eff}({\tilde A}_{eff},\Delta_{12})$ is the
survival rate for the electron neutrino in the effective two-flavor system with the matter
effect ${\tilde A}_{eff}$ and mass squared difference $\Delta_{12}$.
This reduction formula has a twofold improvement compared with the leading order formula Eq. n(\ref{leading}); the matter effect $A_{eff}$ is replaced by ${\tilde
A_{eff}}$ and the angle is corrected from $\theta_{13}$ to
$\theta_{13}+\epsilon$ at the production point of the neutrino, $t=0$ \footnote{Though this {\it angle} correction is
pointed out in Ref. \cite{Smi1}, it is neglected because of its
smallness.}.

This next-to-leading formula is valid no matter whether or not the resonance due to
the matter effect occurs.  We can, however, further simplify
this relation if the effective survival rate ${\tilde S}_{2\nu
,eff}({\tilde A}_{eff},\Delta_{12})$ is represented using the jump
probability between the two mass eigenstates at the resonance point
$P_c({\tilde A_{eff}},\theta,\Delta)$, which is almost the same as that for $A_{eff}$. The two-flavor survival rate is represented using
the jump probability as \cite{Par}\footnote{This relation is valid
only if the survival rate is averaged over with respect to the initial
time.  Strictly speaking, we therefore can not use the obtained
simplified reduction formula without averaging over the survival rate
with respect to the initial time.  This is not the case for the
reduction formula obtained in the present paper Eq.(\ref{final}).}
\begin{eqnarray}
{\tilde S}_{2\nu ,eff}
=
\frac{1}{2}
+
\left\{\frac{1}{2}-P_c({\tilde A_{eff}},\theta_{12},\Delta_{12})\right\}
\cos{ 2{\tilde\theta_{m 0}}}\cos{2\theta_{12}}
\label{parke}
\end{eqnarray}
where ${\tilde \theta}_{m0}$ is the mixing angle in the matter at the neutrino
production point,
\begin{eqnarray}
\cos{2{\tilde \theta_{m 0}}}
=
\frac{\Delta_{12}\cos{2\theta_{12}}-{\tilde A_{eff}}(0)}
{\sqrt{(\Delta_{12}\cos{2\theta_{12}}-{\tilde A_{eff}}(0))^2
+(\Delta_{12}\sin{2\theta_{12}})^2}}.
\label{cosm0}
\end{eqnarray}
We say that the resonance is complete in the case that the electron neutrino is produced far above the resonance point. 
In this case, the survival rate is represented by a jump probability as in Eq. (\ref{parke}). 
On the other hand, we say that the resonance is incomplete in the case that the neutrino
is produced too near the resonance point. 
In this case, we cannot represent the survival rate by the jump probability. 
The survival rate for an
electron neutrino produced at a point with infinite matter effect,
$A_{eff}=c_{13}^2 A \rightarrow \infty$, is represented by
\begin{eqnarray}
S_{2\nu ,eff}
=
\frac{1}{2}
-
\left\{\frac{1}{2}-P_c( A_{eff},\theta_{12},\Delta_{12})\right\}
\cos{2\theta_{12}}.
\label{parke2}
\end{eqnarray}
For the case $P_c({\tilde
A_{eff}},\theta_{12},\Delta_{12})\simeq P_c(A_{eff},\theta_{12},\Delta_{12})$,
the survival rate ${\tilde S}_{2\nu ,eff}$ is rewritten as
\begin{eqnarray}
{\tilde S}_{2\nu ,eff}
=
\frac{1}{2}
-
\left(
S_{2\nu ,eff}
-\frac{1}{2}
\right)
\cos{2{\tilde\theta_{m 0}}}.
\label{srelation}
\end{eqnarray}
Substituting this relation into Eq. (\ref{final}), we obtain the
simplified formula
\begin{eqnarray}
S_{3\nu}(\nu_e \rightarrow \nu_e)
=
\cos^{2}{\theta_{13}}\cos^2({\theta_{13}+\epsilon})
\left\{\frac{1}{2}
-
\left(
S_{2\nu ,eff}
-\frac{1}{2}
\right)
\cos{2{\tilde\theta_{m 0}}}
\right\}
+
\sin^{2}{\theta_{13}}\sin^2({\theta_{13}+\epsilon}).\nonumber\\
\label{simple}
\end{eqnarray}

In some cases, the survival rate $S_{2\nu ,eff}$ is known exactly, and the
simplified formula can be represented in completely analytic form. For example, when
the matter density distribution is of the exponential type, $A=A_0 \exp
(-\frac{r}{r_0})$, which we adopt in the numerical analysis in the next
section, the survival rate is represented analytically using the
jump probability,
\begin{eqnarray}
P_c(A,\theta,\Delta)
=
\frac{\exp (2 \pi \Delta r_0 \cos\theta^2)-1}
{\exp (2 \pi \Delta r_0)-1}.
\label{expjump}
\end{eqnarray}
Using Eqs. (\ref{aeffdef}),(\ref{final}),(\ref{cosm0}),(\ref{parke2}),(\ref{srelation}),(\ref{expjump}), the survival rate for the three-flavor neutrino oscillation $S_{3\nu}$ is represented
analytically under the condition $P_c({\tilde A_{eff}},\theta_{12},\Delta_{12})\simeq
P_c(A_{eff},\theta_{12},\Delta_{12})$, in this case. We notice that this
simplification is valid only in the case that the survival rate can be
represented by the jump probability.  It cannot be used in the case of
incomplete resonance as is seen from the numerical analysis in the
next subsection.

\subsection{Numerical confirmation of the validity of the
next-to-leading order reduction formula}\label{onenum}

In the present subsection, we numerically examine the precision of the next-to-leading
order reduction formula.  As an
example, we use an exponential type electron density distribution
$N(r)=245 N_A \exp({-10.54\frac{r}{R_{\odot}}})$, which is a good
approximation for the solar neutrino propagation in most regions \cite{Bah}\footnote{This
approximation is not so good near the core and the surface of the sun.
The density around the core causes the error of the reduction formula
through the angle, $\epsilon$.  Since the real density is smaller than
the value obtained by the exponential approximation, the real error of
the reduction formula for the sun is expected to be smaller than the
value obtained in the present numerical calculation. }.  Here, $N_A$
is Avogadro's number.  We consider the mass squared differences
$10^{-10}<\delta m_{12}^{2}<10^{-4} [\textrm{eV}^2],\
1.5\times 10^{-3}<\delta m_{13}^{2}<4\times
10^{-3} [\textrm{eV}^2]$ and the whole range of mixing angles $0 < \theta_{12}
< \frac{\pi}{2},\ 0 < \theta_{13} < \frac{\pi}{2}$.  We average
over the survival rate with respect to the final time, and do not do it
with respect to the initial time, since we do not use averaging
in deriving Eq. (\ref{final}).


Since we have a vast parameter region, we do not investigate the whole
parameter region.  Instead, we investigate the parameter region where
the error of the reduction formula is expected to be large, since we
would like to be conservative concerning the precision of the
reduction formula.  We first fix the values of the mass squared
differences as $\delta m_{12}^{2}=10^{-4} [\textrm{eV}^2]$ and $\delta
m_{13}^{2}=\ 1.5\times 10^{-3} [\textrm{eV}^2]$, where the hierarchy between
the mass differences is minimum in the region we consider and the
error is expected to be maximally enhanced.

\begin{figure}
\unitlength=1cm
\begin{picture}(17,18)
\unitlength=1mm
\centerline{
\includegraphics[scale=0.8]{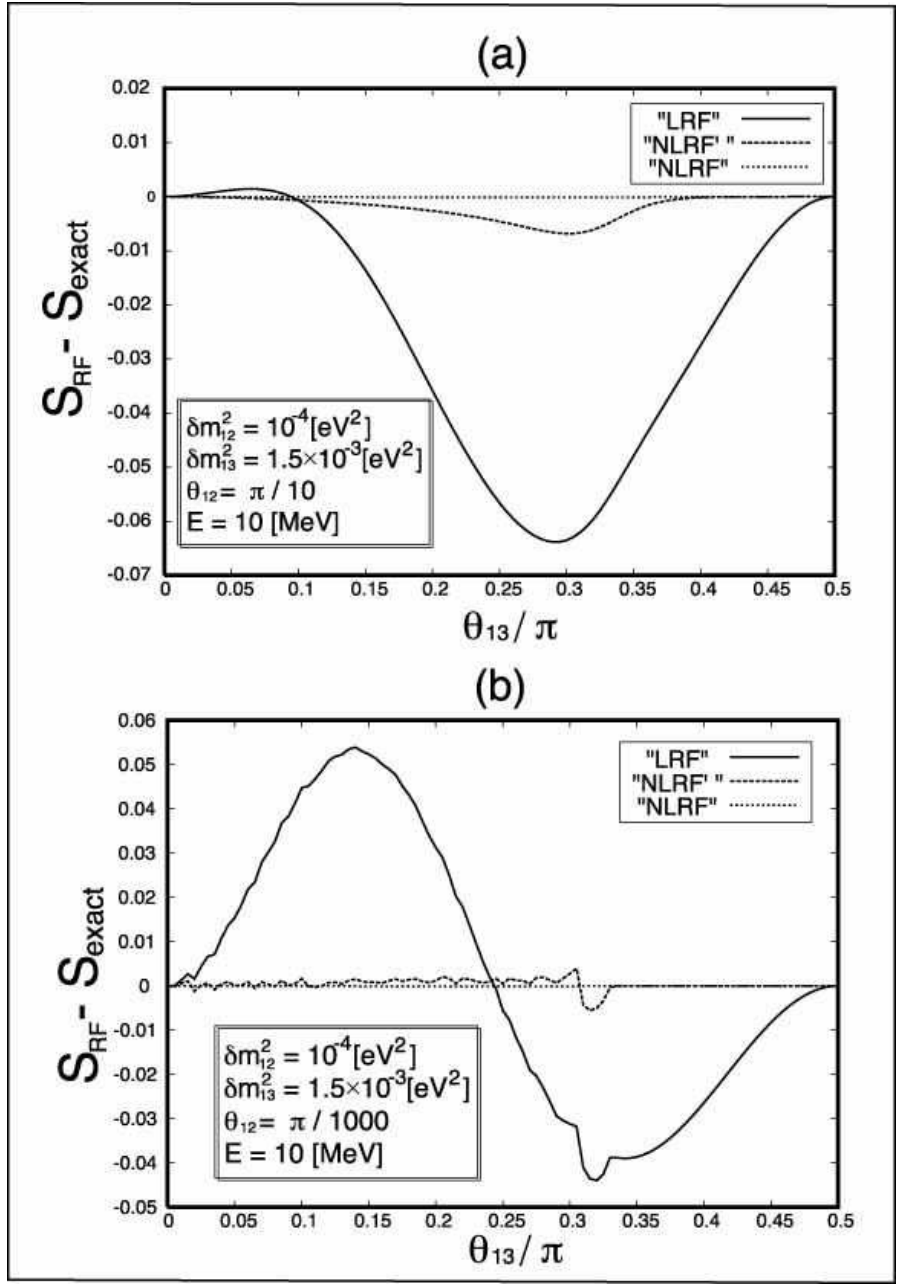}
} 
\end{picture}
\caption{The error of the reduction formula is shown for the
parameters written in the figure as a function of $\theta_{13}$ in
(a).  The result using the leading formula Eq. (\ref{leading}), denoted by "LRF", has the largest error
of about 0.07. The next-to-leading order formula Eq. (\ref{final})
without the replacement $A_{eff}\rightarrow {\tilde A_{eff}}$, denoted by "NLRF$'$", has an error of about 0.01.  The next-to-leading order formula
Eq. (\ref{final}) with ${\tilde A}_{eff}$, denoted by "NLRF", has the smallest error of order $10^{-4}$. Larger errors arise for larger $\sin 2\theta_{13}$. (b) The same (a) for smaller value of $\theta_{12}$.The magnitudes of each error are almost the same as those of
(a). They arise, however, at relatively small values of $\theta_{13}$.
\label{theta3}}
\end{figure}

We show the error of the reduction formula, the deviation from the
exact result $S_{RF}-S_{exact}$, as a function of $\theta_{13}$ in
Fig. \ref{theta3}(a).  The result derived from the leading formula
Eq. (\ref{leading}) is known to have the largest error of about
0.07. The next-to-leading order formula Eq. (\ref{final}) without the
replacement $A_{eff}\rightarrow {\tilde A_{eff}}$ has an error of
about 0.01.  The next-to-leading order formula Eq. (\ref{final}) with
the replacement $A_{eff}\rightarrow {\tilde A_{eff}}$ has the smallest
error, of order $10^{-4}$.  
We show the same figure for smaller value of $\theta_{12}$
in Fig. \ref{theta3}(b).  The magnitudes of each error are almost the
same as those of Fig. \ref{theta3}(a).  Theyarise, however, at relatively
small value of $\theta_{13}$ in this case.  Generally, the errors are
small for small $\sin 2\theta_{13}$ as is expected from the fact that
the angle correction $\epsilon$ is proportional to $\sin 2\theta_{13}$
(Eq. ({\ref{epsilon}})) and the correction to the matter effect is
proportional to $\sin^2 2\theta_{13}$ (Eq. (\ref{aefflim})).

\begin{figure}
\unitlength=1cm
\begin{picture}(17,17)
\unitlength=1mm
\centerline{
\includegraphics[scale=0.8]{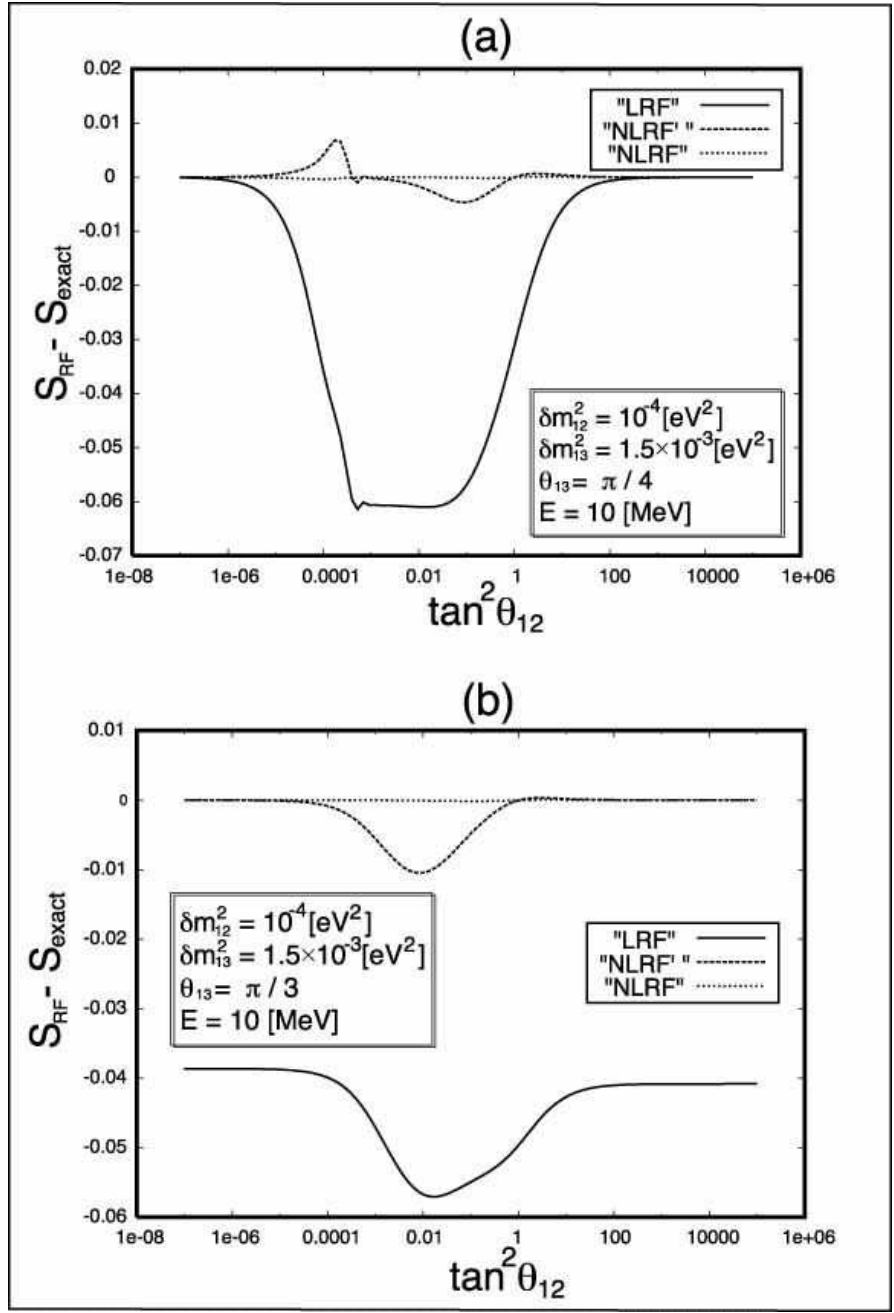} 
} 
\end{picture}
\caption{(a) The error of the reduction formula for the parameters
written in the figure as a function of $\theta_{12}$.  The leading
formula Eq. (\ref{leading}) has the largest error of about 0.06.  The
next-to-leading order reduction formula Eq. (\ref{final}) without the
replacement $A_{eff}\rightarrow {\tilde A_{eff}}$ has error around
0.01.  The next-to-leading order reduction formula Eq. (\ref{final})
with the replacement $A_{eff}\rightarrow {\tilde A_{eff}}$ has an error of only order $10^{-4}$. Although the error of the
leading order approximation vanishes for extremely large or small
values of $\theta_{12}$, this is accidental one. (b) Same as (a) for $\theta_{13}=\frac{\pi}{3}$.
Although the magnitudes of the errors are almost the same as
in (a) there remains a finite error for extremely large or
small values of $\theta_{12}$. (b) shows the result for the same parameter set except for $\theta_{13}=\frac{\pi}{3}$.
\label{theta1}}
\end{figure}

We show the results of the error of the reduction formula 
as a function of $\theta_{12}$ for $\sin^2 2\theta_{13}=1$ in
Fig. \ref{theta1}(a), where the error is expected to be large. 
The leading order formula Eq. (\ref{leading}) has
the largest error of about 0.06.  The next-to-leading order reduction
formula Eq. (\ref{final}) without the replacement $A_{eff}\rightarrow
{\tilde A_{eff}}$ has an error around 0.01.  The next-to-leading order
reduction formula Eq. (\ref{final}) with the replacement
$A_{eff}\rightarrow {\tilde A_{eff}}$ has an error of order 
$10^{-4}$. Although the error of the leading order approximation
vanishes for extremely large or small values of $\theta_{12}$, this is
accidental. Since the survival rate ${\tilde S}_{2\nu,eff}$ is apparently unity for
extremely small or large $\theta_{12}$, the error comes from the
angle correction $\epsilon$ in this region. The correction of $S_{3\nu}$ due to the mixing angle
correction is estimated to be
\begin{eqnarray}
\frac{\sin^2 2\theta_{13} A(0)}{2\Delta_{13}}
(s_{13}^2-c_{13}^2{\tilde S}_{2\nu ,eff})
\label{angleerror}
\end{eqnarray}
from Eqs. (\ref{epsilon}),(\ref{final}).  For
$\theta_{13}=\frac{\pi}{4}$ and ${\tilde S}_{2\nu ,eff}=1$, this
quantity accidentally vanishes.  This error therefore remains for other
values of $\theta_{13}$ as shown in Fig. \ref{theta1}(b), which is the
same figure as Fig. \ref{theta1}(a) except for $\theta_{13}=\frac{\pi}{3}$.

\begin{figure}
\unitlength=1cm
\begin{picture}(17,18)
\unitlength=1mm
\centerline{
\includegraphics[scale=0.8]{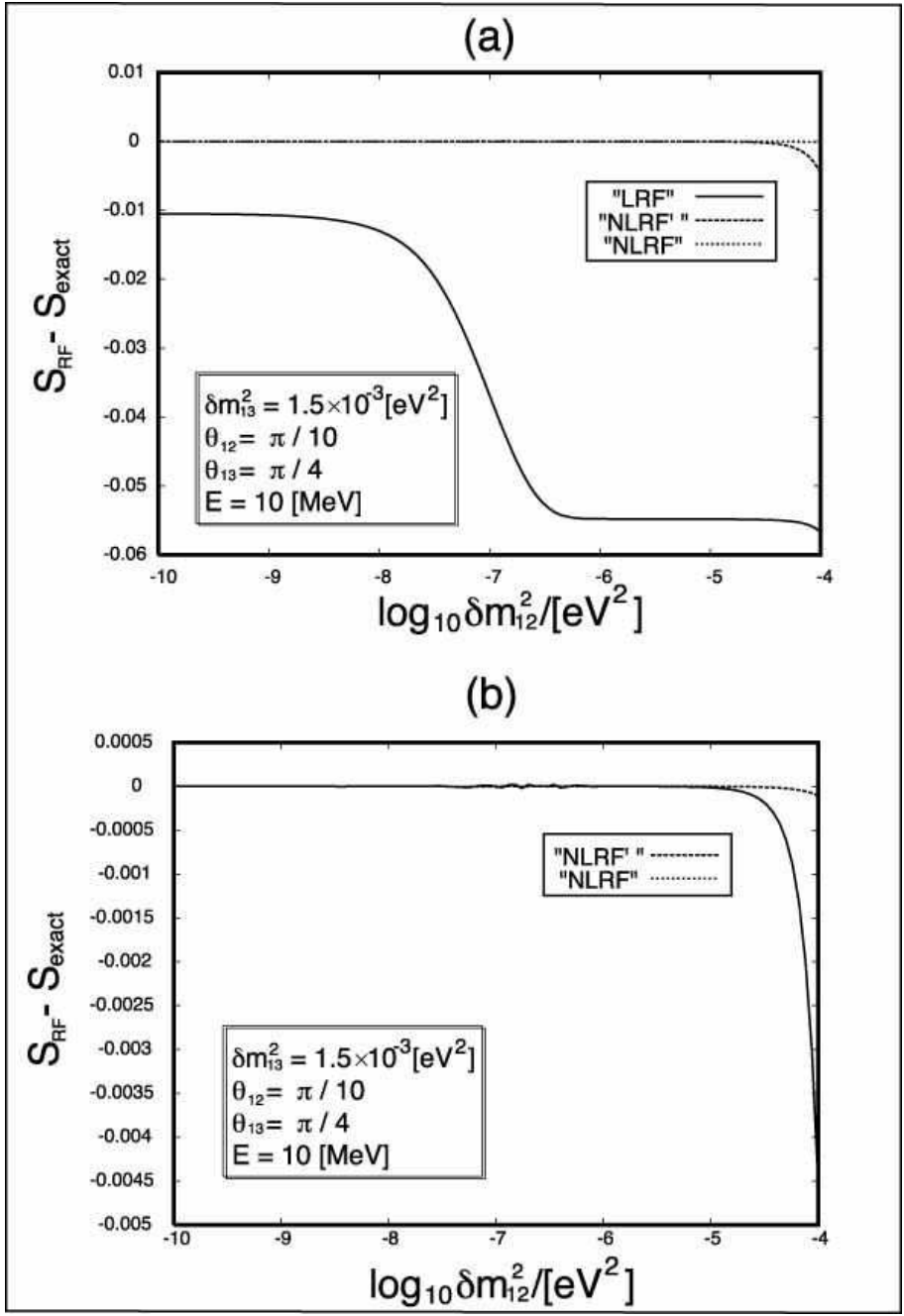}
} 
\end{picture}
\caption{The error of the reduction formula is shown for the parameters
written in the figure as a function of $\delta m^{2}_{12}$. 
(a) and (b) are the same except for the scale of the vertical axis. 
The leading formula Eq. (\ref{leading}) has the largest error of about 0.05.  The
other lines, which have errors less than 0.01, are the
next-to-leading order formula Eq. (\ref{final}) with and without the
replacement $A_{eff}\rightarrow {\tilde A_{eff}}$. For the leading
formula, the error remains even for small $\delta m_{12}^2$, although it tends to vanish for
the next-to-leading order formula.
\label{m12}}
\end{figure}

 We next confirm that the error tends to be smaller for a larger
hierarchy, i.e., smaller $\delta m_{12}^2$ and larger $\delta
m_{13}^2$, as is suggested by the fact that the errors are handled by
the relative importance of the neglected off-diagonal elements
compared with the dominant matrix element $\Delta_{13}$,
i.e., $\frac{A(0)\sin 2\theta_{13}}{2\Delta_{13}}$ for the leading
order formula and $\frac{\Delta_{12}\sin
2\theta_{12}\epsilon}{2\Delta_{13}}\sim \frac{\Delta_{12}\sin
2\theta_{12}A(0)\sin 2\theta_{13}}{4\Delta^{2}_{13}}$ for the
next-to-leading order formula.  We first show the $\delta m_{12}^2$
dependence of the error in Fig. \ref{m12}.  The leading formula
Eq. (\ref{leading}) has the largest error as in the above cases.  There
remains a finite error, even for extremely small values of $\delta
m_{12}^2$.  
The errors for the next-to-leading order approximations tend to vanish
for smaller values of $\delta m_{12}^2$.  This is because the error is
handled by $\frac{\Delta_{12}\sin 2\theta_{12}A(0)\sin
2\theta_{13}}{4\Delta^{2}_{13}}$, and suggests that the correction to
the matter effect Eq. (\ref{aeffdef}) is negligible in this case.
Therefore, the correction to the matter effect will be negligible for
the LOW and VO solutions of the solar neutrino oscillation.

\begin{figure}
\unitlength=1cm
\begin{picture}(22,10)
\unitlength=10mm
\centerline{
\includegraphics[scale=1.0]{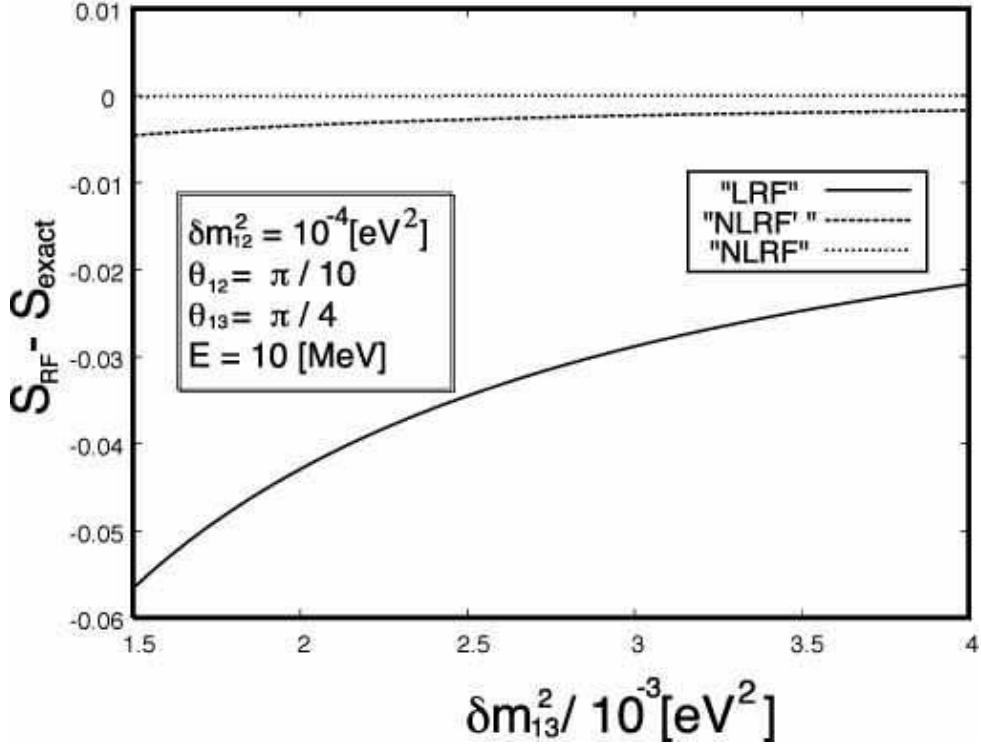}
} 
\end{picture}
\caption{The error of the reduction formula is shown for the parameters
written in the figure as a function of $\delta m^{2}_{13}$.  The leading
formula Eq. (\ref{leading}) has the largest error of about 0.06.  The
other lines, which have errors less than 0.01, are the
next-to-leading order formula Eq. (\ref{final}), with and without the
replacement $A_{eff}\rightarrow {\tilde A_{eff}}$. All errors tend
to be reduced for larger $\delta m_{13}^2$.
\label{m13}}
\end{figure}

We next show the dependence of the error on the mass
difference $\delta m_{13}^2$ in Fig. \ref{m13}.  The leading formula
Eq. (\ref{leading}) again has the largest error.  The
error tends to be reduced for larger values of $\delta m_{13}^2$, as is
expected.

All of the above numerical results strongly suggest that the reduction
formula at the leading order potentially has an error around 0.1.  This
corresponds to the fact that the error due to the neglect of the mixing angle
correction $\epsilon(0)$ is estimated as
\begin{eqnarray}
\frac{\sin^2 2\theta_{13} A(0)}{2\Delta_{13}}
(s_{13}^2-c_{13}^2{\tilde S}_{2\nu ,eff})
\lesssim
0.1,
\label{error}
\end{eqnarray}
from Eqs. (\ref{epsilon}),(\ref{final}).  The numerical calculations
also suggest that the next-to-leading order correction drastically
improves the reduction formula and the error is reduced to be of order 
$10^{-4}$.  These results indicate that the leading order reduction
formula is pretty good for a rough estimation of the allowed
parameter region and the next-to-leading order formula is necessary for
its precise determination. A detailed analysis of the allowed parameter region from the 
solar neutrino data is not with in the scope of the present paper.

\begin{figure}
\unitlength=1cm
\begin{picture}(22,10)
\unitlength=1mm
\centerline{
\includegraphics[scale=1.0]{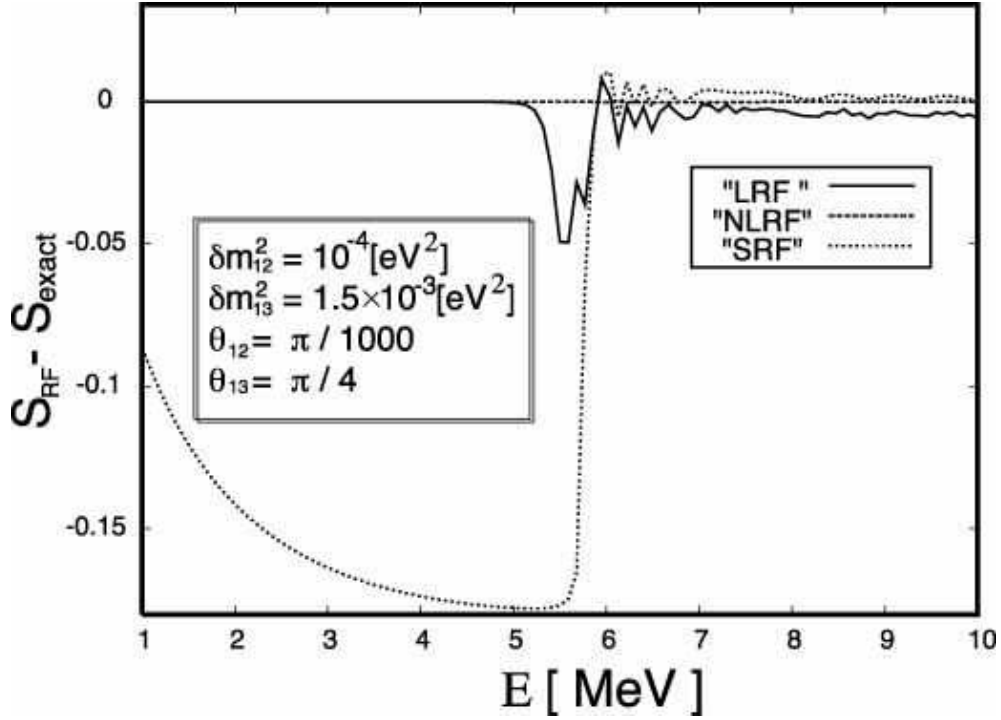}
} 
\end{picture}
\caption{The error of the reduction formula is shown for the
parameters written in the figure as a function of the neutrino energy {\it E}.
The error of the simplified formula is very large for smaller values of
the neutrino energies. On the other hand, the leading order reduction formula has an
error around 0.06. The next-to-leading order reduction formula has an
error of order $10^{-4}$.
\label{energy}}
\end{figure}

Finally, we examine how precise the simplified reduction formula
is \footnote{ Strictly speaking, we can not use the simplified
reduction formula without averaging over the survival rate with
respect to the initial time.  The error due to this, however, seems to
be very small numerically.  }.  Since this formula is derived under
the condition that the survival rate can be expressed by the jump
probability, it will not be precise when the resonance is
incomplete.  For the case of the above example, $N(r)=245 N_A
\exp({-10.54\frac{r}{R_{\odot}}})$, the resonance is incomplete for
neutrinos which have smaller energies $\sim 1$ [MeV].  The simplified
reduction formula therefore is not expected to be valid in this case.
On the other hand, the simplified formula is expected to be precise
for neutrinos that have larger energies $\sim 10$ [MeV], because in
this case the resonance is expected to occur almost completely.  We
show in Fig. \ref{energy} the error of the reduction formula as a
function of the neutrino energy {\it E}.  One can observe that the
simplified formula is not valid for smaller energies of the
neutrino from this figure.  The errors both of the leading and the
next-to-leading order reduction formulas are also shown in the figure.
According to the above result, we learn that we should not use the
simplified formula without ensuring that the resonance is complete, although
the formula is attractive because of its simplicity.

\begin{figure}
\unitlength=1cm
\begin{picture}(17,18)
\unitlength=1mm
\centerline{
\includegraphics[scale=0.8]{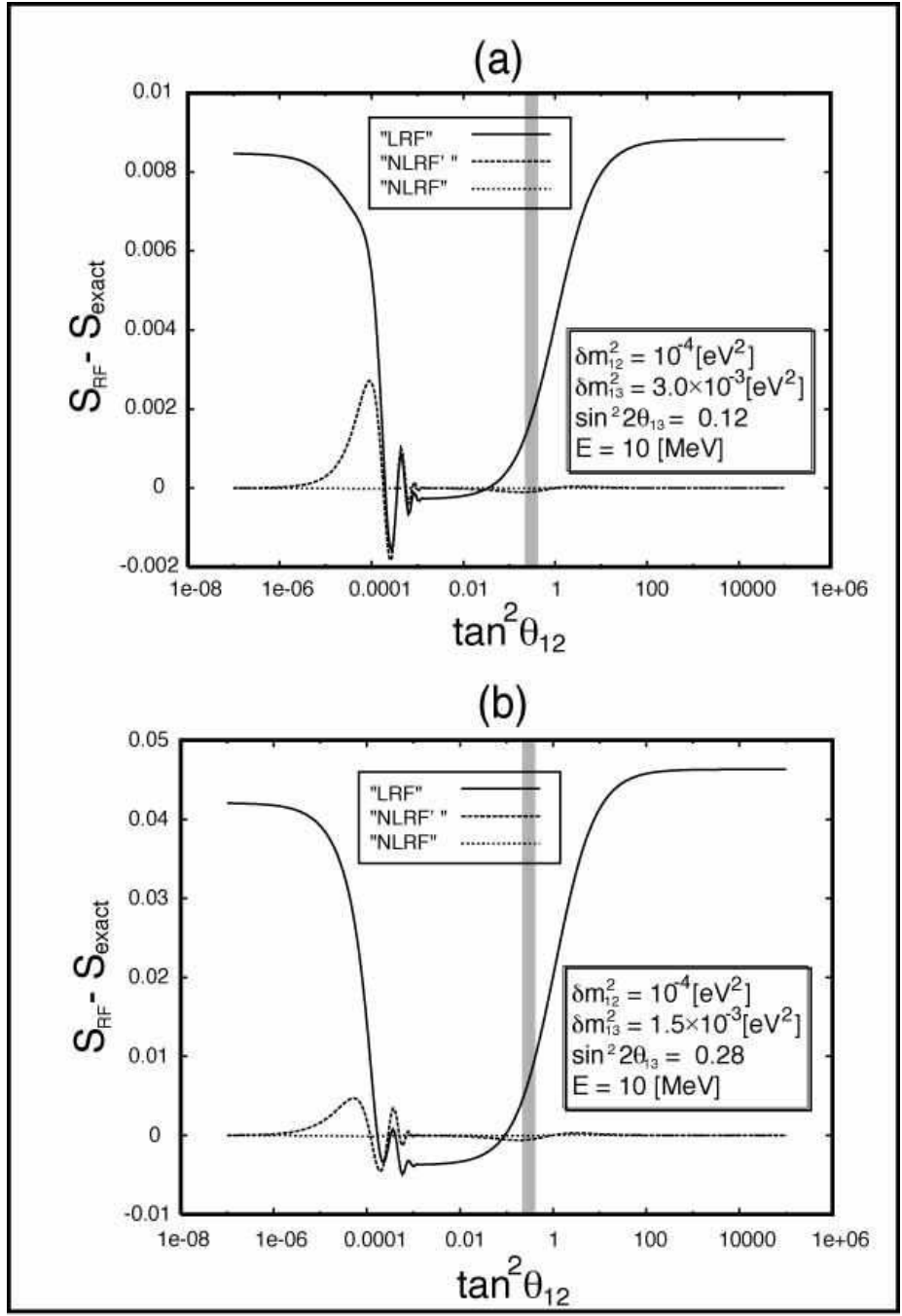}
} 
\end{picture}
\caption{The error of the reduction formula for  the realistic parameters (a) $\delta m_{13}^2=3.0 \times 10^{-3} [\textrm{eV}^2], \sin^2{2\theta _{13}=0.12}$ and (b) $ \delta m_{13}^2=1.5 \times 10^{-3} [\textrm{eV}^2], \sin^2{2\theta _{13}=0.28}$.  The error of the leading formula Eq. (\ref{leading}) is shown as "LRF".  The
errors of the
next-to-leading order formula Eq. (\ref{final}) without and with the
replacement $A_{eff}\rightarrow {\tilde A_{eff}}$ are shown as "NLRF$'$" and "NRF", respectively. The allowed region from the solar neutrino data is shaded.  The error for the case (b) is found to be larger than that of (a), and the errors are rather small for the shaded region in both cases.}
\label{realistic}
\end{figure}

So far we have not restricted the values of the mixing angles, since our aim was
to conservatively clarify the precision of the next-to-leading order reduction
formula.
However, there exist meaningful upper bounds on the mixing angle
$\theta _{13}$ from the reactor experiments CHOOZ \cite{CHOOZ} and Palo Verde \cite{Palo}.
The smaller the mixing angle $\theta _{13}$ becomes, the better the precision of reduction
formulas is expected to be. Thus we have calculated the errors of
the reduction formulas for  $\mbox{sin}^{2} \ 2\theta _{13} = 0.12$, the upper
bound corresponding to the best fit value $\delta m_{13}^{2} = 3\times
10^{-3}~[\textrm{eV}^{2}] $ to account for the
atmospheric neutrino data \cite{SuperK}. The results are shown in Fig. \ref{realistic}
(a), as functions of the remaining mixing angle $\mbox{tan} ^{2}
\theta _{12} $. Also shown by a shaded area is the region allowed at 90\%
C.L. by the LMA MSW solution \cite{SNO}. We learn from this figure that the
leading order formula has an error up to 1\% or so for some values of
$\mbox{tan} ^{2} \theta _{12} $, while the error of the next-to-leading order
formula  Eq.(\ref{final}) is essentially negligible. The error of the leading-order
formula is, however, less than 0.2\% if we remain in the shaded region. In
Fig. \ref{realistic}(b), we have also shown the errors of the reduction formulas for
$\mbox{sin}^{2}2\theta _{13}=0.28$, the upper bound corresponding to
the value $\delta m_{13}^{2} = 1.5 \times 10^{-3}~[\textrm{eV}^{2}] $, the lowest
mass squared difference to account for the atmospheric neutrino data
\cite{SuperK} at 90\% C.L. We now learn that the error of the leading order formula is
enhanced by both smaller $\delta m_{13}^{2}$ and larger
$\theta _{13}$; the error reaches
5\% or so.  Even for the shaded region, the error can be up to 1\%.

From these analyses, we can say that for the values of $\theta _{13}$
implied by the reactor experiments \cite{CHOOZ} the error of the
leading order formula can be rather small, while that of the next-to-leading order 
formula is completely negligible. If we further impose the condition
suggested by the LMA MSW solution, the error is even smaller, i.e., at most
1\%. Such an error, however, will be problematic for precision tests of neutrino experiments, such as the (indirect) search for CP violation \cite{Joe,Xin,Bra,Smi3}, which require the precise determination of the size of the
unitarity triangle with an accuracy of a few percent \cite{Smi3}, and therefore a
better precision of the reduction formula itself.
We hope that the next-to-leading order reduction formula proposed here, being a simple formula to use, will be useful for study of the precision tests of neutrino experiments.

\section{Reduction formula in the case of two resonances}

We derive a reduction formula when $\nu _{e}$ experiences two successive
resonances in Sec. \ref{twoder}.  The expression we derive is slightly different from previously proposed ones and is applicable for wider situations.  This formula is
relevant for investigation into the supernova neutrino data or
the hypothetial very high energy solar neutrino
data due to the annihilation of WIMPs \cite{Kuo2,Dut,Smi4,Fog4,Fog5,Gou}.  We then verify the validity
of the reduction formula using the numerical calculation in Sec \ref{twonum}.

\subsection{Derivation of the reduction formula in the case of two resonances
\label{twoder}}

Suppose $\nu _{e}$ is produced at time 0 and detected at time $t$, going through "higher" and "lower" resonances, 
caused by matching the matter effect with $\Delta _{13}$ and $\Delta _{12}$, respectively.
We divide the time interval into [0,$t_{M}$] and [$t_{M}$,$t$] ($0 < t_{M} < t$), where the conditions $\Delta _{12} \ll \Delta _{13},A$ 
and $\Delta _{12}, A \ll \Delta _{13}$ are met, i.e., the higher and lower resonances are operative.
The intermediate time $t_{M}$ is chosen so that $\Delta _{12} \ll A(t_{M}) \ll \Delta _{13}$.

We first consider the time range  where the higher resonance occurs, i.e.,
$\Delta_{12} \ll \Delta_{13},\ A(t)$.  A convenient base to describe this region
is
\begin{eqnarray}
\omega \equiv U \phi,\hspace{1cm}
U
=
\left(
\begin{array}{ccc}
0 & -1 & 0 \\
1 & 0  &0\\
0 & 0  & 1 \\
\end{array}
\right)
V^{\dag}_{23},
\end{eqnarray}
where the time evolution equation Eq. (\ref{prope}) can be cast into
\begin{eqnarray}
i\frac{d\omega}{dt}
&=&
\left\{
UV
\left(
\begin{array}{ccc}
0 &  &  \\
 & \Delta_{12} &  \\
 &  & \Delta_{13} \\
\end{array}
\right)
V^{\dagger}U^{\dagger}
+
U
\left(
\begin{array}{ccc}
A(t) &  &  \\
 & 0 &  \\
 &  & 0 \\
\end{array}
\right)
U^{\dag}
\right\}
\omega\\
&=&
\left(
\begin{array}{ccc}
c_{12}^2\Delta _{12}
& -s_{12}c_{12}c_{13}\Delta _{12}
&s_{12}c_{12}s_{13}\Delta _{12}\\
-s_{12}c_{12}c_{13}\Delta _{12}
&s^{2}_{12}c^{2}_{13}\Delta _{12}+s^{2}_{13}\Delta _{13}+A(t)
&s_{13}c_{13}\Delta _{13}-s^{2}_{12}s_{13}c_{13}\Delta _{12} \\
s_{12}c_{12}s_{13}\Delta _{12} 
&s_{13}c_{13}\Delta _{13}-s^2_{12}s_{13}c_{13}\Delta _{12}
&s_{12}^{2}s_{13}^{2}\Delta _{12}+c_{13}^2\Delta _{13} \\
\end{array}
\right)
\omega\\
&\simeq&
\left(
\begin{array}{ccc}
\cos 2\theta _{12}\Delta_{12}& 0&0\\
0&s_{13}^2(\Delta _{13}-s_{12}^2\Delta _{12})+A(t)&s_{13}c_{13}(\Delta _{13}-s_{12}^2\Delta _{12})\\
0 &s_{13}c_{13}(\Delta _{13}-s_{12}^2\Delta _{12})
&c_{13}^2(\Delta _{13}-s_{12}^2\Delta _{12})\\
\end{array}
\right)
\omega\nonumber\\
&&+s_{12}^2\Delta _{12}I\omega\\
&\equiv &
\left(
\begin{array}{cc}
\cos 2\theta_{12}\Delta _{12}& {\bf 0}^{\bf T}\\
{\bf 0} & {\bf M}_H(t) \\
\end{array}
\right)
\omega.
\label{propomega}
\end{eqnarray}
Here, we neglect the term proportional to the unit matrix $I$.

The resolvent matrix in this base $R^H$ is
\begin{eqnarray}
R_H(t,0)
\simeq
\left(
\begin{array}{cc}
e^{-i\cos 2\theta_{12}\Delta_{12} t } & {\bf 0}^{\bf T}\\
{\bf 0} & {\bf R}_{2\times 2}^H(t,0)\\
\end{array}
\right),
\end{eqnarray}
where ${\bf R}_{2\times 2}^H$ is the resolvent matrix for the
effective two-flavor neutrino system with matter effect $A(t)$, mass
squared difference $\Delta_{13}-s_{12}^2\Delta_{12}$, and mixing
angle $\theta_{13}$.

We next consider the time range, [$t_{M}$,$t$], where the lower resonance occurs, i.e., $\Delta_{12},
A(t) \ll \Delta_{13}$.  A convenient base in this region is the time
dependent base $\kappa (t)\equiv Z(t) \phi$, by which the Hamiltonian for fixed $t$ is diagonalized, 
\begin{eqnarray}
Z(t)
\left\{
V
\left(
\begin{array}{ccc}
0 &  &  \\
 & \Delta_{12} &  \\
 &  & \Delta_{13} \\
\end{array}
\right)
V^{\dagger}
+
\left(
\begin{array}{ccc}
A(t) &  &  \\
 & 0 &  \\
 &  & 0 \\
\end{array}
\right)
\right\}
Z^{\dag}(t)
=
\left(
\begin{array}{ccc}
k_1(t) & 0 & 0 \\
0 & k_2(t)  &0\\
0 & 0  & k_3(t) \\
\end{array}
\right)\hspace{1cm} (k_1<k_2<k_3).\nonumber\\
&&
\end{eqnarray}
Because of the hierarchy $\Delta _{12}, A \ll \Delta _{13}$, the Hamiltonian and therefore the resolvent are approximately 
block diagonalized:
\begin{eqnarray}
R_L(t,t_M)
&\simeq &
\left(
\begin{array}{cc}
{\bf R}_{2\times 2}^L(t,t_M) & {\bf 0}\\
{\bf 0}^{\bf T} & e^{-i\int_{t_M}^{t} k_{3}(t')dt'}\\
\end{array}
\right).
\end{eqnarray}
Since the lower resonance is nearly complete, i.e., $\Delta _{12}\ll A(t_{M})\ll \Delta_{13}$, the $2\times 2$ resolvent matrix can be generally written as
\begin{eqnarray}
{\bf R}_{2\times 2}^L(t,t_M)
&\simeq&
\left(
\begin{array}{cc}
1 & 0\\
0 & e^{-i\Delta_{12}(t-t_L)}\\
\end{array}
\right)
\left(
\begin{array}{cc}
\sqrt{1-P_L}e^{i\alpha} & \sqrt{P_L}e^{i\beta}\\
\sqrt{P_L}e^{i\gamma} & -\sqrt{1-P_L}e^{-i(\alpha-\beta-\gamma)} \\
\end{array}
\right),\label{res2L}
\end{eqnarray}
where $t_L$ is some time far after the lower resonance, i.e., $A(t_{L}) \ll \Delta _{12}$, and $P_L$ is the
jump probability between adiabatic states with respect to the lower resonance.

Using these resolvent matrices, the survival rate of $\nu_{e}$ is given as
\begin{eqnarray}
S
&=&
\left|
\left(
\begin{array}{ccc}
1& 0& 0
\end{array}
\right)
Z^{\dag}(t)
\left(
\begin{array}{cc}
{\bf R}_{2\times 2}^L(t,t_M) &{\bf 0}\\
{\bf 0}^{\bf T} & \exp({-i\int_{t_M}^{t} k_3(t') dt'})\\
\end{array}
\right)
Z(t_M)
\right.
\nonumber\\
&&\hspace*{4cm}
\left.
U^{\dagger}
\left(
\begin{array}{cc}
e^{-i\cos 2\theta _{12} \Delta_{12} t_M } & {\bf 0}^{\bf T}\\
{\bf 0} & {\bf R}_{2\times 2}^H(t_M,0)\\
\end{array}
\right)U
\left(
\begin{array}{c}
1\\
0\\
0\\
\end{array}
\right)
\right|^2.
\label{2resS}
\end{eqnarray}
Here, the matrix $Z(t_M)U^{\dagger}$ is approximately
written as
\begin{eqnarray}
Z(t_M)U^{\dagger}
\simeq
\left(
\begin{array}{cc}
1 & {\bf 0}^{\bf T}\\
{\bf 0} & {\bf V}_{3m}^{\dagger}(t_M)\\
\end{array}
\right)
\simeq
\left(
\begin{array}{ccc}
1&0&0\\
0&c_{13}&-s_{13}\\
0&s_{13}&c_{13}\\
\end{array}
\right),\label{tramat}
\end{eqnarray}
where ${\bf V}_{3m}^{\dagger}(t)$ is the matrix that diagonalizes
the $2\times 2$ mass matrix ${\bf M}_H$ as
\begin{eqnarray}
{\bf V}_{3m}^{\dagger}(t)
{\bf M}_H(t)
{\bf V}_{3m}(t)
=
\left(
\begin{array}{cc}
k_2(t)  &0\\
0  & k_3(t) \\
\end{array}
\right).
\end{eqnarray}
From Eq. (\ref{2resS}) and Eq. (\ref{tramat}) and $U(1,0,0)^{T}=(0,1,0)^{T}$,
\begin{eqnarray}
S
\simeq
\left|
\left(
\begin{array}{ccc}
1& 0& 0
\end{array}
\right)
Z^{\dag}(t)
\left(
\begin{array}{cc}
{\bf R}_{2\times 2}^L(t,t_M) &{\bf 0}\\
{\bf 0}^{T} & \exp({-i\int_{t_M}^{t} k_3(t') dt'})\\
\end{array}
\right)
\left(
\begin{array}{c}
0\\
{\bf V}_{3m}^{\dagger}(t_M){\bf R}_{2\times 2}^H(t_M,0)
\left(
\begin{array}{c}
1\\
0\\
\end{array}
\right)
\\
\end{array}
\right)
\right|^2.\nonumber\\
\end{eqnarray}
In order to express the survival rate in terms of the two-flavor survival rate,
we set
\begin{eqnarray}
\left(
\begin{array}{c}
a\\
b\\
\end{array}
\right)
=
{\bf V}_{3m}^{\dagger}(t_M){\bf R}_{2\times 2}^H(t_M,0)
\left(
\begin{array}{c}
1\\
0\\
\end{array}
\right).
\end{eqnarray}
The survival rate for the two-flavor system with respect to the higher
resonance is expressed by these quantities as
\begin{eqnarray}
S_{2\times 2}^H
&=&
\left|
\left(
\begin{array}{cc}
1& 0
\end{array}
\right)
{\bf R}_{2\times 2}^H(t,0)
\left(
\begin{array}{c}
1\\
0\\
\end{array}
\right)
\right|^2\\
&=&
\left|
\left(
\begin{array}{cc}
1& 0
\end{array}
\right)
{\bf V}_{3m}(t){\bf V}_{3m}^{\dagger}(t)
{\bf R}_{2\times 2}^H(t,0)
\left(
\begin{array}{c}
1\\
0\\
\end{array}
\right)
\right|^2\\
&\simeq&
\left|
\left(
\begin{array}{cc}
1& 0
\end{array}
\right)
{\bf V}_{3m}(t)
\left(
\begin{array}{cc}
e^{-i\int_{t_M}^{t} k_2(t') dt'} & 0\\
0 & e^{-i\int_{t_M}^{t} k_3(t') dt'}\\
\end{array}
\right)
\left(
\begin{array}{c}
a\\
b\\
\end{array}
\right)
\right|^2\\
&\longrightarrow&
|a|^2 \cos^2{\theta_{13}}+|b|^2\sin^2{\theta_{13}}.
\label{highsurvival}
\end{eqnarray}
In Eq. (\ref{highsurvival}), we averaged with respect to the final time $t$ and used
\begin{eqnarray*}
{\bf V}_{3m}(t)\simeq
{\bf V}_{3m}(t_{M})\simeq
\left(
\begin{array}{cc}
c_{13}&s_{13}\\
-s_{13}&c_{13}\\
\end{array}
\right).
\end{eqnarray*}
Using the unitarity condition $|a|^2 + |b|^2 =1$, we get
\begin{eqnarray}
|a|^2
=
\frac{S_{2\times 2}^H-\sin^2{\theta_{13}}}{\cos{2 \theta_{13}}},\ \ 
|b|^2
=
\frac{\cos^2{\theta_{13}}-S_{2\times 2}^H}{\cos{2 \theta_{13}}} .\label{ab}
\end{eqnarray}
The survival rate is thus expressed as
\begin{eqnarray}
S
&=&
\left|
\left(
\begin{array}{ccc}
1& 0& 0
\end{array}
\right)
Z^{\dag}(t)
\left(
\begin{array}{c}
{\bf R}_{2\times 2}^L(t,t_M)
\left(
\begin{array}{c}
0\\
a\\
\end{array}
\right)\\
b\exp{(-i\int_{t_M}^{t} k_3(t') dt')}
\end{array}
\right)
\right|^2\\
&\longrightarrow &
|a|^2 \cos^2{\theta_{13}} \{P_L \cos^2{\theta_{12}}+(1-P_L)\sin^2{\theta_{12}}\}
+
|b|^2 \sin^2{\theta_{13}}.
\label{prefinal}
\end{eqnarray}
In Eq. (\ref{prefinal}), we averaged the survival rate with respect to the
final time and used $Z^{\dag}(t)=V$.  Using Eq. (\ref{ab}), we get the reduction formula, which is our final result,
\begin{eqnarray}
S
=
S_{2\times 2, inf}^L 
\frac{S_{2\times 2}^H-\sin^2{\theta_{13}}}{\cos{2\theta_{13}}}
\cos^2{\theta_{13}}
+
\frac{\cos^2{\theta_{13}}-S_{2\times 2}^H}{\cos{2 \theta_{13}}}
\sin^2{\theta_{13}}.\label{twofinal}
\end{eqnarray}
Here, $S_{2\times 2, inf}^L (=P_{L}\cos ^{2}\theta _{12}+(1-P_{L})\sin ^{2}\theta _{12})$ is the two-flavor survival rate for the
lower resonance in the case that the electron neutrino is produced at
a point with $A\rightarrow \infty$.  We verify the validity of this formula by a numerical method in the
next subsection.

This reduction formula coincides with the reduction formula obtained
in Ref. \cite{Gou}:
\begin{eqnarray}
S
&=&
P_2^H \cos^2{\theta_{13}} S_{2\times 2, inf}^L
+
(1-P_2^H ) \sin^2{\theta_{13}},
\label{Gred}\\
&&
\left(P_2^H
=
\frac{1}{2}
+
(\frac{1}{2}-P_H)
\cos{2 \theta_{13m}}
\right),\nonumber
\end{eqnarray}
provided the higher resonance is complete and the two-flavor survival rate $S_{2\times 2}^H$ can be
written in terms of the jump probability $P_H$ as
\begin{eqnarray}
S_{2\times 2}^H
=
\frac{1}{2}
+
(\frac{1}{2}-P_H)
\cos{2 \theta_{13m}}\cos{2 \theta_{13}}.
\end{eqnarray}
This reduction formula Eq. (\ref{Gred}) is the same as the reduction
formula Eq. (\ref{leading2}) obtained in Ref. \cite{Kuo}, 
if $\theta _{13}$ is set to be $\frac{\pi}{2}$ and $S^{L}_{2\times 2, inf}$ is further expressed by the jump probability $P_{L}$ as 
$S^{L}_{2\times 2, inf}=\frac{1}{2}-\left(\frac{1}{2}-P_{L}\right)\cos 2\theta _{12}$.
Although the reduction formulas Eq. (\ref{twofinal}) and Eq. (\ref{Gred}) are
the same when the higher resonance is complete, they are different when the higher resonance is incomplete \footnote{Since the higher
resonance is complete at the parameters in Ref \cite{Gou}, the use of
the reduction formula Eq. (\ref{Gred}) is valid there as is shown in
Sec. \ref{twonum}}. We also confirm the difference numerically in the next
subsection.  

\subsection{Numerical confirmation of the validity of the reduction
formula for the case of two resonances}
\label{twonum}

In the present subsection, we examine the precision of the reduction
formula for the case of two resonances Eq. (\ref{twofinal}) using a
numerical calculation.  As a typical example, we use the same electron
density distribution $N(r)=245 N_A \exp({-10.54\frac{r}{R_{\odot}}})$
as in Sec. \ref{onenum}.  Since we consider the case where two
resonances occur, we assume that the energy of the produced electron
neutrino is very high compared to that of the solar neutrino, $E > 100$
[MeV]. 
This situation is quite similar to that considered in
Ref. \cite{Gou}. To get conservative results, we take the mass squared differences as $\delta
m_{12}^{2}=10^{-4} [\textrm{eV}^2]$ and $\delta m_{13}^{2}=\ 1.5\times 10^{-3}
[\textrm{eV}^2]$, where the hierarchy between the mass squared differences is
mildest in the experimentally allowed region and the error is expected to be enhanced maximally.

\begin{figure}
\unitlength=1cm
\begin{picture}(22,20)
\unitlength=1mm
\centerline{
\includegraphics[scale=0.8]{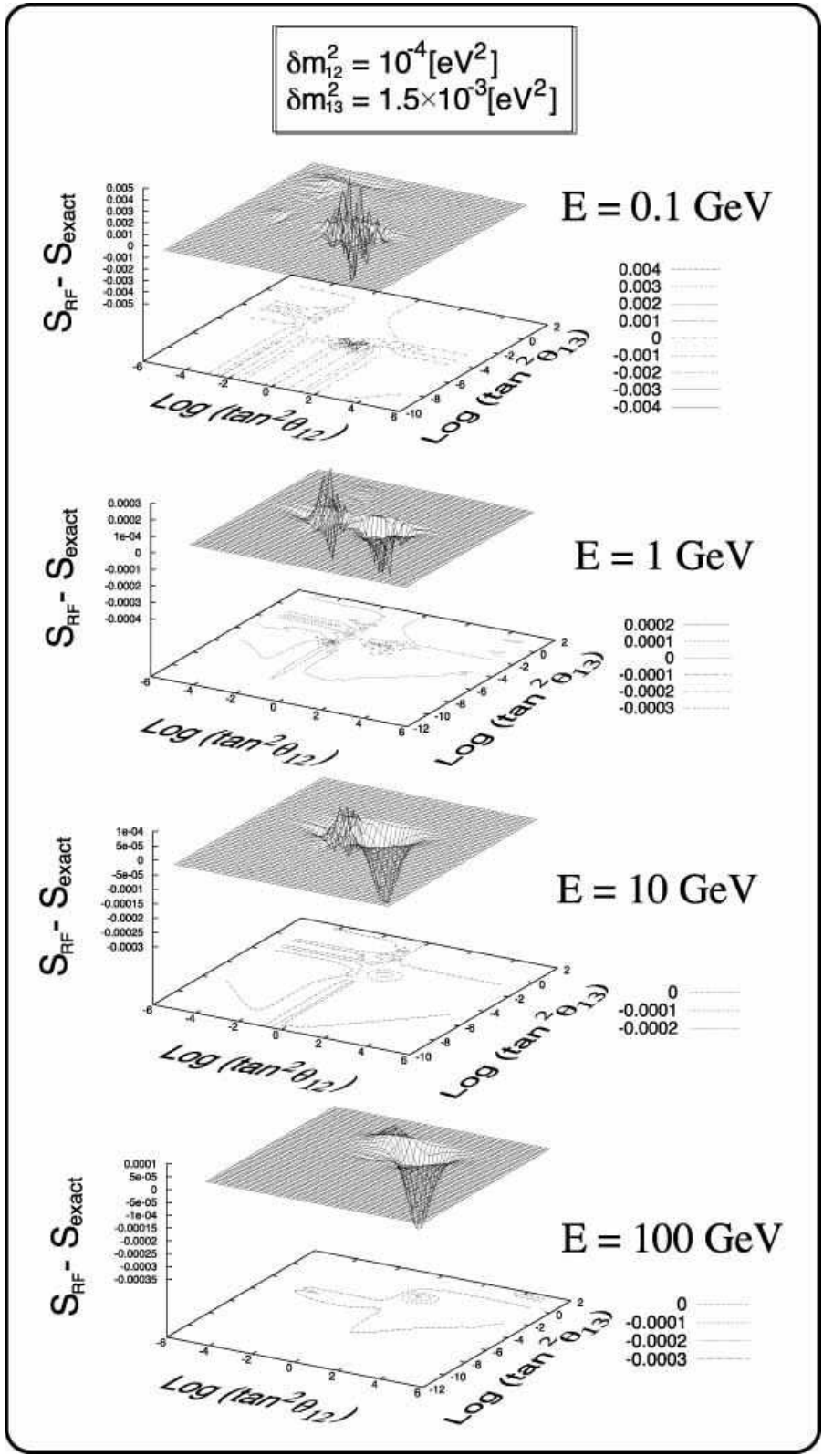}
} 
\end{picture}
\caption{The error of the reduction formula Eq. (\ref{twofinal}) is
shown for the parameters written in the figure as a function of
$\theta_{12}$ and $\theta_{13}$ for various values of the neutrino
energy $E=0.1,1,10,100$ [GeV].  The largest error occurs in the case of the
lowest energy $E=0.1$ [GeV].  Even in this case, the error has
magnitude less than 0.005.
\label{twocon}}
\end{figure}

We show the error of the reduction formula Eq. (\ref{twofinal}) in
Fig. \ref{twocon} for various values of the electron energy $E = 0.1,
1,10,100$ [GeV].  We observe the largest error for the lowest energy 
$E = 0.1$ [GeV]. Even in this case, however, the error is less than 0.005. 
For higher energy neutrinos, the reduction formula is more
accurate.  This reduction formula is apparently accurate enough for
the investigation into supernova neutrinos and the hypothetical O(GeV) solar
neutrinos \cite{Kuo2,Dut,Smi4,Fog4,Fog5,Gou}.

\begin{figure}
\unitlength=1cm
\begin{picture}(22,10)
\unitlength=1mm
\centerline{
\includegraphics[scale=1.0]{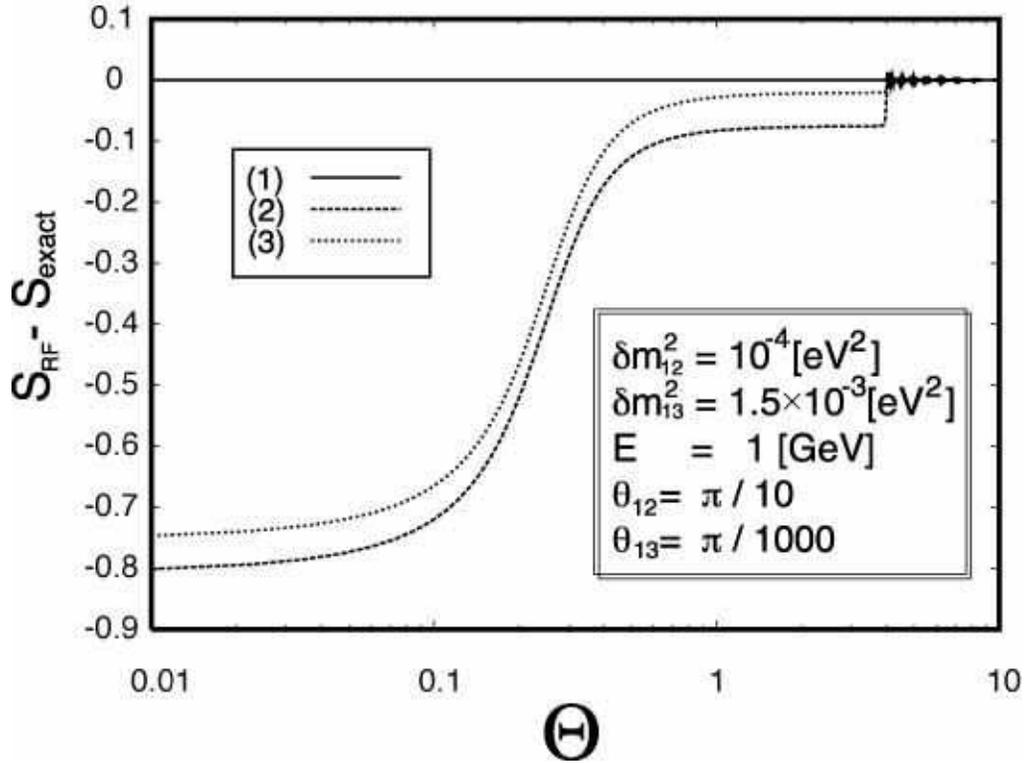}
} 
\end{picture}
\caption{The errors of the reduction formulas Eq. (\ref{twofinal}),
Eq. (\ref{Gred}), and Eq. (\ref{leading2}) as a function of the initial
electron density.  The horizontal axis is $\Theta$ which is defined as  
$N(r)=245 \Theta N_A \exp({-10.54\frac{r}{R_{\odot}}})$. 
The lines (1), (2) and (3) correspond to Eqs. (\ref{twofinal}), (\ref{leading2}), and (\ref{Gred}), respectively. 
The reduction
formula Eq. (\ref{twofinal}) is extremely accurate for all values of
$\Theta$. While the reduction formula Eq. (\ref{Gred}) is better than
Eq. (\ref{leading2}), neither is accurate for smaller values of $\Theta$.
\label{twocom}}
\end{figure}

We finally compare the reduction formula obtained here
[Eq. (\ref{twofinal})] with that obtained in Ref. \cite{Gou}.  We compare the
errors as a function of the initial electron density $N(0)=245\Theta N_{A}$ in
Fig. \ref{twocom}. The reduction formula Eq. (\ref{twofinal}) is
extremely accurate for all values of $\Theta$. While the reduction
formula Eq. (\ref{Gred}) is better than Eq. (\ref{leading2}), it is not
accurate for smaller values of $\Theta$ where the higher resonance
becomes incomplete. The error, however, is 
small enough in the parameter range considered in Ref. \cite{Gou} for their
purpose.  This result shows that the reduction formula
Eq. (\ref{twofinal}) can be safely used even in the region where the
higher resonance is incomplete.

\section{Summary}
\label{sum}

In the present paper, we derive the next-to-leading order reduction formula
for the $\nu _{e}$ survival rate Eq. (\ref{final}) from the three-flavor neutrino oscillation to the two-flavor one in the case when there is only one resonance, as in the ordinary solar
neutrino oscillation.  Together with an analytic argument, we numerically verify the accuracy of the reduction formula, leaving the mixing angles free for generality.  While we find that the leading order reduction formula
Eq. (\ref{leading}) is accurate enough for a rough
estimation, the next-to-leading order reduction formula is extremely
accurate and adequate for precision tests of neutrino oscillations.  
Next, we study the accuracy of the reduction formulas in a realistic case, i.e., taking into account the current upper bound on $\theta _{13}$. 
We find that the largest error of the leading order reduction formula is about 1\% or so while the error of the next-to-leading order reduction formula is negligible. 
We thus point out that this precise next-to-leading order formula will be useful for precision tests of neutrino oscillations, for example the (indirect) study of CP violation
\cite{Joe,Xin,Bra,Smi3}. 
We also verify the accuracy of the reduction formula written using the jump probability.
This formula is accurate when the resonance is complete, i.e., for high energy neutrinos, although it
is not valid when the resonance is incomplete, i.e., for low energy
neutrinos. 

We also derive the reduction formula Eq. (\ref{twofinal}) in the case of two resonances as in the oscillations of the supernova neutrinos and the hypothetical high energy O(GeV) solar
neutrinos due to the annihilation of WIMPs.  We numerically verify that
it is quite accurate and applicable for any parameter region.  We then compare it to the
reduction formulas obtained in Refs. \cite{Kuo,Gou}. Although the previously obtained formulas are valid only
in case the higher resonance is complete, the formula obtained here is valid not only for the complete case but also for the incomplete case.

\begin{acknowledgments}
We thanks Y. Saruki for his valuable input. 
One of the authors (K.O.) thanks JSPS Grant No. 4834 for financial support. The work of C.S.L. was supported in part by a Grant-in-Aid for scientific Research of the Ministry of Education, Science, and Culture, Grant No. 80201870.
\end{acknowledgments}

\end{document}